\documentclass[12pt]{article}
\usepackage{graphicx}

\newcommand\pubnumber{}
\newcommand\pubdate{\today}

\def\Title#1{\begin{center} {\Large #1 } \end{center}}
\def\Author#1{\begin{center}{ \sc #1} \end{center}}
\def\Address#1{\begin{center}{ \it #1} \end{center}}

\newcommand\pubblock{\rightline{\begin{tabular}{l} \pubnumber\\
         \pubdate  \end{tabular}}}
\newenvironment{Abstract}{\begin{quotation}  }{\end{quotation}}
\newenvironment{Presented}{\begin{quotation} \begin{center} 
             PRESENTED AT\end{center}\bigskip 
      \begin{center}\begin{large}}{\end{large}\end{center} \end{quotation}}
\def\Acknowledgements{\bigskip  \bigskip \begin{center} \begin{large}
             \bf ACKNOWLEDGEMENTS \end{large}\end{center}}

\newlength{\imgwidth}
\def\beq{\begin{equation}}
\def\eeq#1{\label{#1}\end{equation}}
\def\eeqn{\end{equation}}

\def\beqa{\begin{eqnarray}}
\def\eeqa#1{\label{#1}\end{eqnarray}}
\def\eeqan{\end{eqnarray}}

\let\bar=\overbar

\def\L{{\cal L}}

\def\Dslash{\not{\hbox{\kern-4pt $D$}}}
\def\dslash{\not{\hbox{\kern-2pt $\del$}}}

\def\msb{{\bar{\ssstyle M \kern -1pt S}}}

\begin{document}
\begin{titlepage}
\pubblock

\vfill
\Title{Rescattering effects in jet--gap--jet processes}
\vfill
\renewcommand*{\thefootnote}{$\ast$}
\Author{Rafa{\l} Staszewski$^{\ a,}$\footnote{\texttt{rafal.staszewski@ifj.edu.pl}}, Izabela Babiarz$^{\ b}$, Antoni Szczurek$^{\ a,b}$}
\Address{$^a$ Institute of Nuclear Physics Polish Academy of Sciences\\Radzikowskiego 152, 31-342 Krak{\'o}w, Poland}
\Address{$^b$ Faculty of Mathematics and Natural Sciences, University of Rzesz{\'o}w\\Pigonia 1, 35-310 Rzesz{\'o}w, Poland}
\vfill
\begin{Abstract}
  We discuss the process of colour-singlet parton--parton scattering, which in hadron--hadron collision can lead to production of jet events, where a rapidity gap between the jets is present. The dynamics of the process is calculated within LL BFKL approximation. Using \textsc{Pythia} MC generator, hadronic shapes of jet--gap--jet events are studied, in particular the distributions of the rapidity gap size resulting from the jet formation processes. Using \textsc{Pythia}'s modelling of multiple parton interactions, the rescattering effects are simulated. It is shown how these effects influence the gap size distributions. The kinematic dependence of the gap survival probability is discussed.
\end{Abstract}
\vfill
\begin{Presented}
EDS Blois 2017, Prague, \\ Czech Republic, June 26-30, 2017
\end{Presented}
\vfill
\end{titlepage}
\def\thefootnote{\fnsymbol{footnote}}
\setcounter{footnote}{0}

\section{Introduction}

One of the most discussed issues in diffractive physics is the problem of rescattering effects.
They originate from the fact that hadrons are finite-size objects of composite structure.
This makes hadron--hadron interactions complex.
In particular, if a diffractive exchange present in a hadron--hadron collision is accompanied by an additional non-diffractive interaction, the overall event would not have a diffractive signature.
This leads to a suppression of diffractive cross sections.

Theoretical calculations often treat the rescattering corrections as a single number, \textit{i.e.} a factor by which the diffractive cross sections are reduced.
This factor is also known as the gap survival probability and various calculations exist for its value at different energies and for different processes, see \textit{e.g.} \cite{Khoze:2000wk}.
However, such a description is only approximate, since the suppression can depend also on the kinematics of the given process.
Therefore, the discussed effect may affect not only the integrated cross sections, but also various kinematic distributions.

In this paper, the diffractive process of the jet--gap--jet production is considered.
The recattering corrections are simulated using the multi-parton interaction (MPI) framework available in \textsc{Pythia} MC event generator \cite{Sjostrand:2014zea}.

\section{Jet--gap--jet processes}

The jet--gap--jet process is a type of jet production in which the interacting partons, which later form the jets, interact by a colour-singlet exchange, see Fig.~\ref{fig:jgj_diagram}.
The lack of colour transfer between the interacting partons leads to a suppressed radiation in the rapidity region between the jets.
This is illustrated in Fig.~\ref{fig:jgj_vs_nondiff}, where the particle density per unit of rapidity is presented for diffractive (jet--gap--jet) and non-diffractive jets.
The results were obtained with hadronisation simulated in \textsc{Pythia} for events with fixed parton-level kinematics --  all events were $gg\to gg$ interactions with gluons scattered at arbitrarily chosen $y = \pm3$ and $p_T=50$\ GeV. 
The experimental studies of this process are based on the signature of a rapidity gap between the jets \cite{D0_jgapj}.

\newlength{\localheight}
\setlength{\localheight}{28ex}
\begin{figure}[htbp]
  \begin{minipage}[t]{0.8\imgwidth}
    \centering
    \includegraphics[height=\localheight]{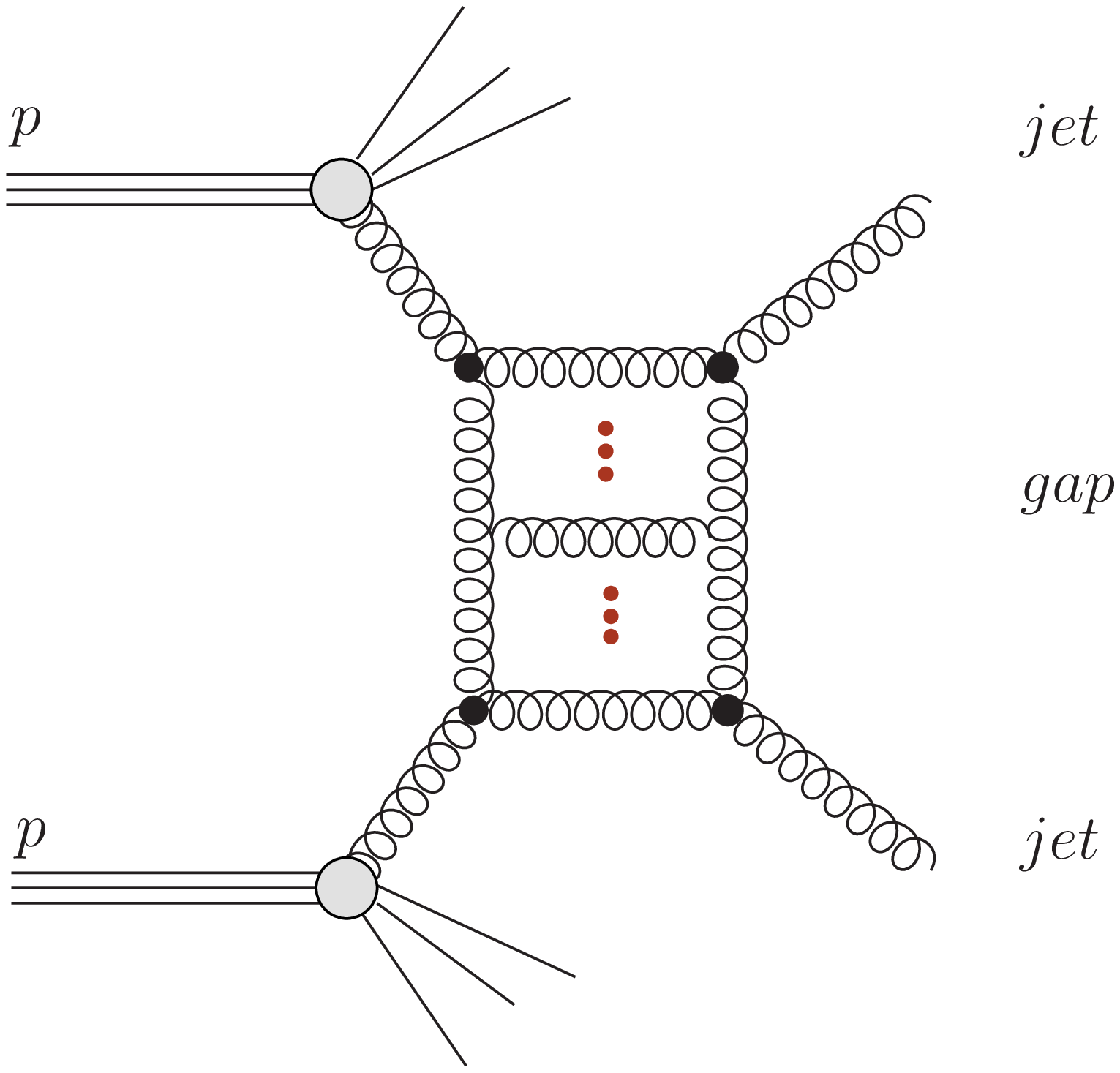}
    \caption{A schematic diagram of the partonic mechanism of the jet--gap--jet process.}
    \label{fig:jgj_diagram}
  \end{minipage}
  \hfill
  \begin{minipage}[t]{1.2\imgwidth}
    \centering
    \includegraphics[height=\localheight]{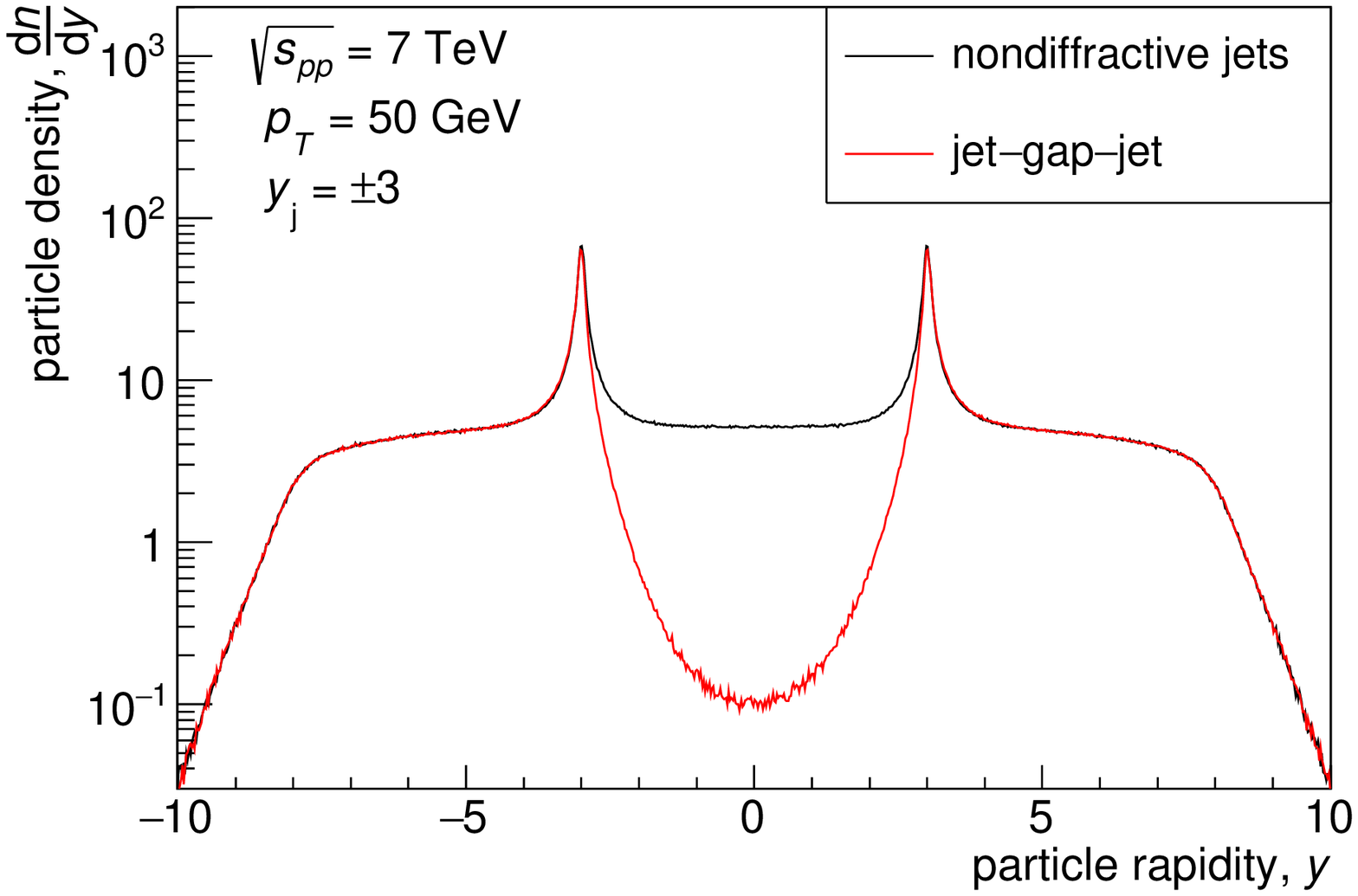}
    \caption{Rapidity distributions of particles produced in
    non-diffractive jet (black) and jet--gap--jet (red), fixed-kinematics events.}
    \label{fig:jgj_vs_nondiff}
  \end{minipage}
\end{figure}

The big advantage of the jet--gap--jet process is that contrary to other diffractive processes the colour singlet is exchanged between partons (not hadrons) and the momentum transfer is large.
These two features allow calculations based on standard parton distributions functions and a hard matrix element, which can be calculated using perturbative methods, see \cite{Mueller:1992pe,Motyka:2001zh}.

\section{Multiple parton interactions}

Since hadrons have complex structure and consist of many partons, it is possible that in a single hadron--hadron collision multiple parton--parton interactions (MPI) occur.

The present work uses the MPI framework provided by the \textsc{Pythia} MC event generator.
The model is based on mini-jet production, for which the divergent behaviour at low transverse momenta is suppressed.
The parameters of this model are chosen such that the predictions agree with the available measurements \cite{event_generator_tunes}.
 
\begin{figure}[htp]
  \centering
  \includegraphics[width=0.49\linewidth]{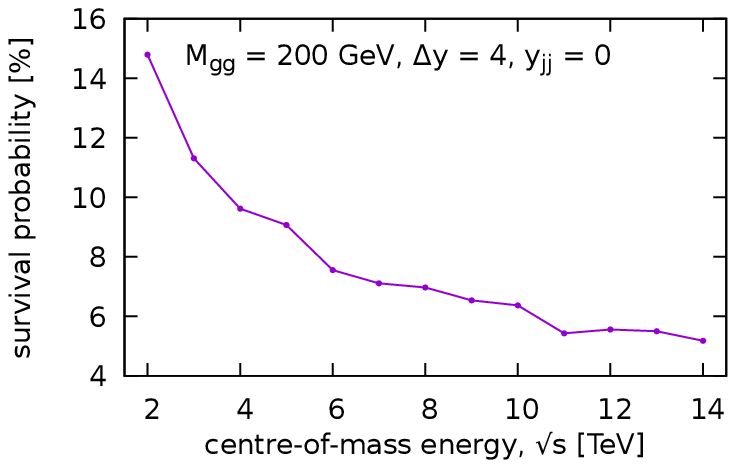}\hfill
  \includegraphics[width=0.49\linewidth]{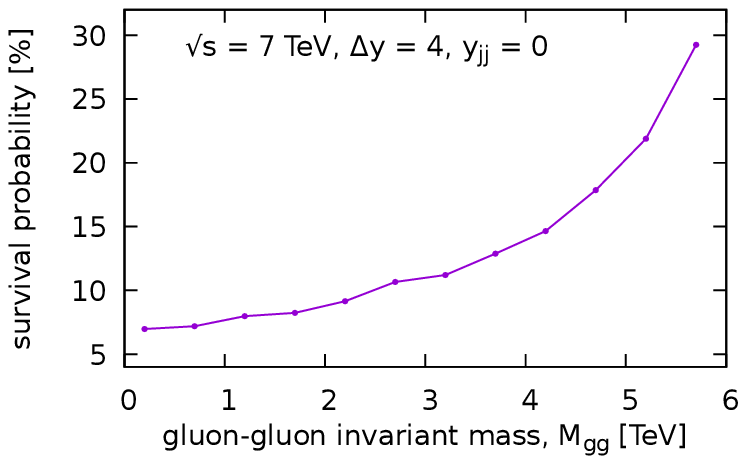}\\
  \caption{Kinematic dependence of gap survival probability as a function of 
  centre-of-mass energy (left) and the hard subprocess invariant mass (right).
  }
  \label{fig:kinematic_dependence}
\end{figure}

Using \textsc{Pythia} MPI framework for generating jet--gap--jet events allows an easy estimation of the rescattering effects and their kinematic dependence.
This can be done by calculating the fraction of events in which no additional MPI took place.
The obtained results are presented in Fig. \ref{fig:kinematic_dependence}.
The results in the left plot agree with the well known decrease of the gap survival probability with the centre-of-mass energy \cite{Khoze:2000wk}.
The right plot shows that at fixed $\sqrt{s}$ the dependence of the actual kinematics of the event can also be significant if jet pairs of very high mass are considered (see also \cite{Babiarz:2017jxc}, where dependence on other kinematic variables is studied).

\begin{figure}[htbp]
  \centering
  \includegraphics[width=\imgwidth]{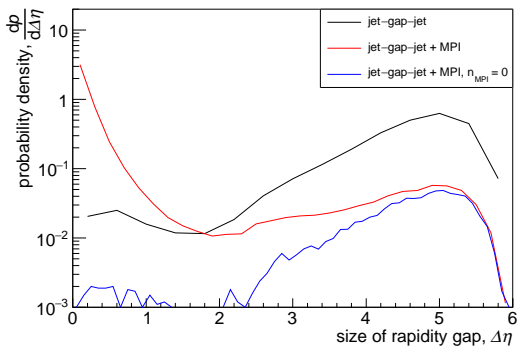}
  \caption{Rapidity gap distributions for jet--gap--jet events: 
  black -- without MPI effects,
  red -- with MPI effects (all events),
  blue -- with MPI effects, only those events in which no MPIs occurred.}
  \label{fig:gap_size}
\end{figure}

Although the above result is closer to the reality than the kinematic-independent gap survival probability, the calculations assume that any MPI would always destroy the diffractive signature of the event.
This turns out not to be a correct assumption.
It is illustrated in Fig. \ref{fig:gap_size}, which presents distributions of rapidity gap size for events simulated without MPI effects, for events simulated with MPI effects and for events simulated with MPI events, but for which no additional MPI occurred.

The results were obtained for events with fixed parton-level kinematics ($y=\pm3$, $p_T=50$~GeV).
The distributions reflect the fluctuations in the hadronisation process and MPI simulation.
The rapidity gap between the scattered gluons is always equal to 6 units.
The fact that the black line in Fig. \ref{fig:gap_size} has a maximum around $\Delta\eta=5$ reflects the typical width of the jets.

The blue line in the plot corresponds to the assumption that the event can be recognised as diffractive only if no additional interactions took place. 
Red line shows the results for the full simulation.
The MPI that accompany the main event produce additional particles in the region between the jets.
This reduces the size of the gap and is responsible for the growth towards $\Delta\eta = 0$.
What is interesting is that even at relatively large gap sizes, \textit{e.g.} $\Delta\eta=3$, there is a difference between the red and the blue curve.
This shows that it is possible that additional interactions do not destroy the gap completely and the event would still have diffractive signature.

\section{Results including jet--gap--jet dynamics}

All above results were obtained using MPI framework for events with fixed parton-level kinematics.
The results presented below were obtained for events generated with jet--gap--jet dynamics based on parton distributions functions and two models for colour-singlet hard exchange: a simple two-gluon exchange, following \cite{BP_book}, and a leading-logarithm BFKL calculation for gluon ladder, following \cite{Chevallier:2009cu}.
In this way, the kinematic dependence discussed above is properly averaged over physical kinematic distributions.
The event generation was performed for $\sqrt{s}=7$~TeV and minimal jet transverse momentum of 20~GeV.

\begin{figure}[htbp]
  \centering
  \includegraphics[width=\imgwidth]{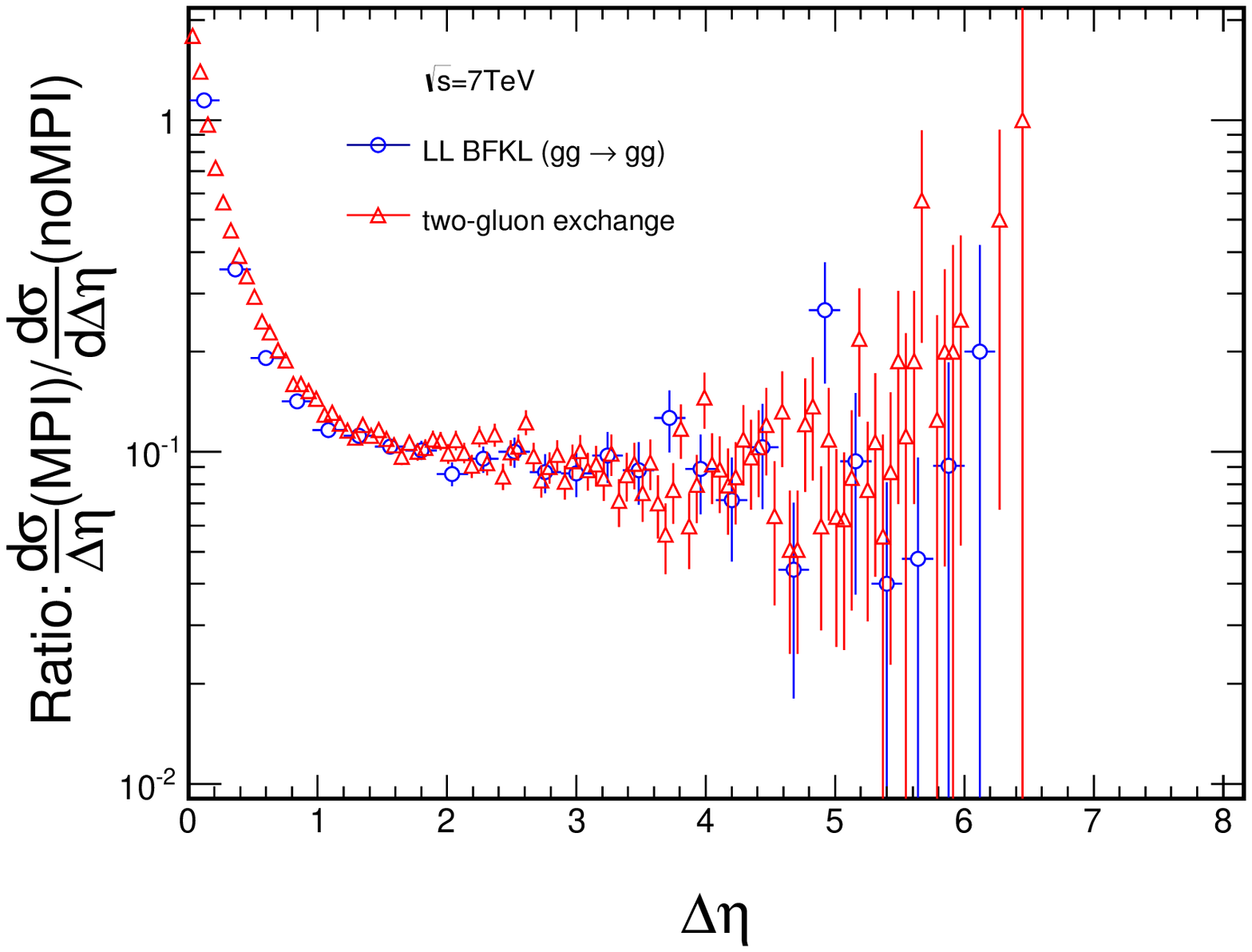}
  \includegraphics[width=\imgwidth]{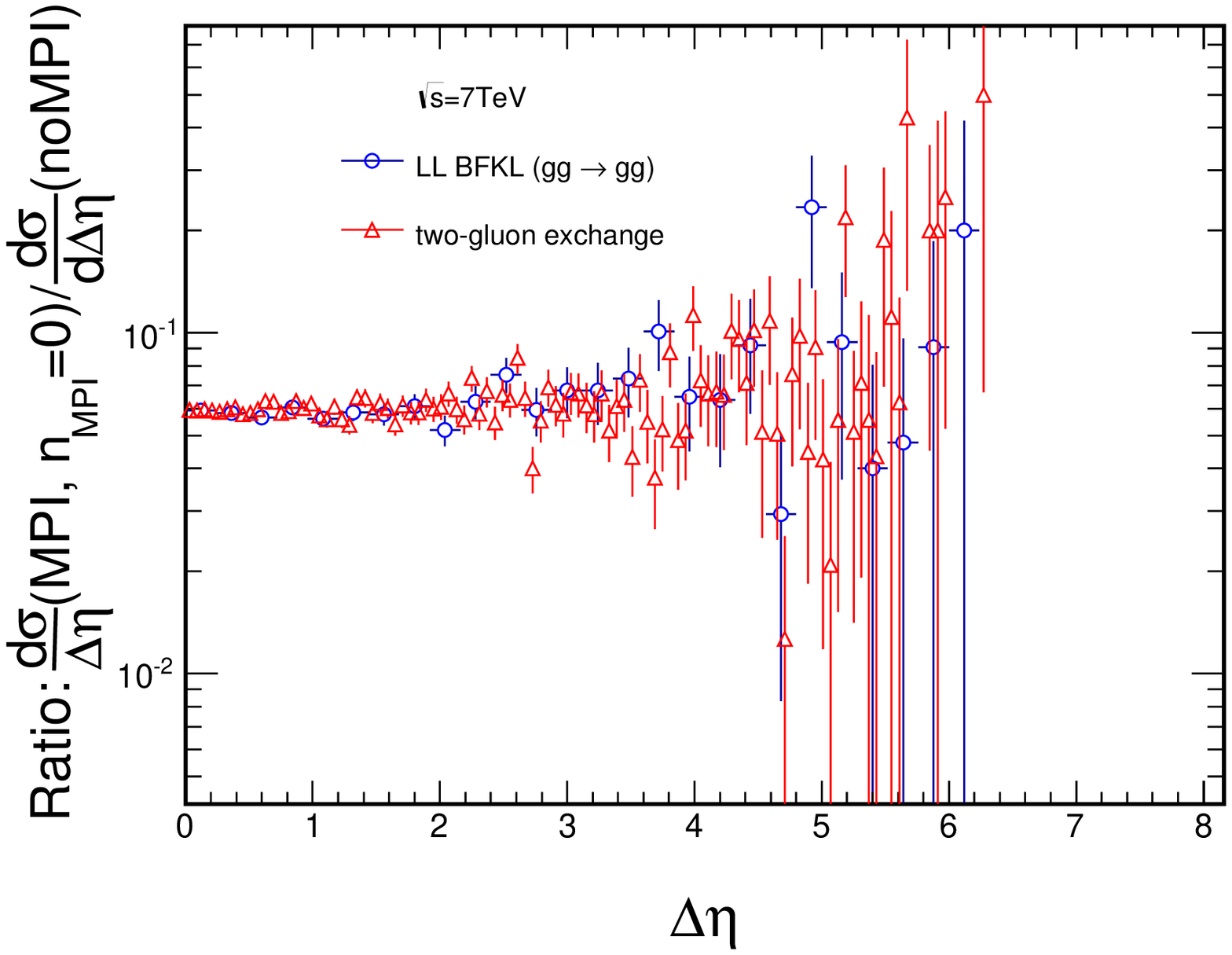}

  \caption{Ratio of rapidity gap distributions for jet--gap--jet events generated with and without MPI effects. 
  Left: nominator contains all events, right: nominator contains only events for which no MPI occurred.}
  \label{fig:ratio}
\end{figure}

Fig. \ref{fig:ratio} presents ratios of rapidity gap distributions for events with MPI effects to events without MPI effects.
For the nominator, the left plots takes all events, while the right plot only those events in which no MPI were present.
The comparison between these two plots quantifies the effects of additional interactions that do not destroy the gap completely. 
Naturally, these effects are very large for small gap sizes.
However, they seem not to disappear even for very large gaps, where the ratio including all events is close to 9\%, while the ratio including only events without additional interactions is about 6\%. The relative difference is therefore non-negligible.

\section{Summary}

The rescattering effects were studied for the jet--gap--jet process.
Using the MPI framework provided by the \textsc{Pythia} event generator the kinematic dependence of the suppression was analysed.
It was observed that this dependence can be very large.
In addition, the possibility that additional interactions do not destroy completely the rapidity gap was studied.
The magnitude of this effect was also observed to be non-negligible.

\Acknowledgements 
This work was supported in part by Polish National Science Centre grant UMO-2015/18/M/ST2/00098.

\end{document}